\documentclass[aps,prl,superscriptaddress,twocolumn,showpacs]{revtex4}
\usepackage{bm}
\usepackage{graphicx}
\usepackage{amsfonts}
\usepackage{amsmath}

\begin{document}

\title{Field-induced $XY$ behavior in the
       $S=\frac12$ antiferromagnet on the square lattice}

\author{Alessandro Cuccoli}
\affiliation{Dipartimento di Fisica dell'Universit\`a di Firenze -
             via G. Sansone 1, I-50019 Sesto Fiorentino (FI), Italy}
\affiliation{Istituto Nazionale per la Fisica della Materia
             - U.d.R. Firenze -
             via G. Sansone 1, I-50019 Sesto Fiorentino (FI), Italy}
\author{Tommaso Roscilde}
\affiliation{Dipartimento di Fisica dell'Universit\`a di Firenze -
             via G. Sansone 1, I-50019 Sesto Fiorentino (FI), Italy}
\affiliation{Istituto Nazionale per la Fisica della Materia
             - U.d.R. Firenze -
             via G. Sansone 1, I-50019 Sesto Fiorentino (FI), Italy}
\author{Ruggero Vaia}
\affiliation{Istituto di Fisica Applicata `Nello Carrara'
             del Consiglio Nazionale delle Ricerche\,-\,via
             Panciatichi~56/30, I-50127 Firenze, Italy}
\affiliation{Istituto Nazionale per la Fisica della Materia
             - U.d.R. Firenze -
             via G. Sansone 1, I-50019 Sesto Fiorentino (FI), Italy}
\author{Paola Verrucchi}
\affiliation{Dipartimento di Fisica dell'Universit\`a di Firenze -
             via G. Sansone 1, I-50019 Sesto Fiorentino (FI), Italy}
\affiliation{Istituto Nazionale per la Fisica della Materia
             - U.d.R. Firenze -
             via G. Sansone 1, I-50019 Sesto Fiorentino (FI), Italy}
\date{\today}

\begin{abstract}
  Making use of the quantum Monte Carlo method based on the worm
  algorithm, we study the thermodynamic behavior of the
  $S\,{=}\,\frac12$ isotropic Heisenberg antiferromagnet on the square
  lattice in a uniform magnetic field varying from very small values
  up to the saturation value.  The field is found to induce a
  Berezinskii-Kosterlitz-Thouless transition at a finite temperature,
  above which a genuine $XY$ behavior in an extended temperature range
  is observed. The phase diagram of the system is drawn, and the
  thermodynamic behavior of the specific heat and of the uniform and
  staggered magnetization is discussed in sight of an experimental
  investigation of the field-induced $XY$ behavior.
\end{abstract}


\maketitle

Field-induced effects in low-dimensional antiferromagnets have been the
subject of renewed interest in the last few years; while on the
experimental side fields of very high intensity have become available,
on the theoretical side the possibility of inducing novel magnetic
phases via application of a strong field have been pointed
out~\cite{Cargese01,TroyerS1998,SchmidTTD2002,Yunoki2002}.

In this paper we consider the two-dimensional quantum Heisenberg
antiferromagnet (2D\,QHAF) in a uniform magnetic field, described by
the Hamiltonian
\begin{equation}
 \hat{\cal H} = \frac{J}{2} \sum_{\bm{i},\bm{d}}
 \hat{\bm S}_{\bm{i}}\cdot\hat{\bm S}_{\bm{i}+\bm{d}}
 - g\mu_{_{\rm B}} H \sum_{\bm{i}}
 \hat{S}^z_{\bm{i}}
\label{e.2DQHAF}
\end{equation}
where ${\bm{i}}=(i_1,i_2)$ runs over the sites of a square lattice,
$\bm{d}$ connects each site to its four nearest neighbors, $J>0$ is the
antiferromagnetic exchange coupling, $H$ is the applied Zeeman field,
and $|S|^2\,{=}\,S(S+1)$. We will hereafter use reduced temperature and
magnetic field, $t\equiv{T/J}$ and $h\equiv{g}\mu_{_{\rm B}}H/(JS)$.

The rich phenomenology~\cite{deJongh1990} of the model is ruled by the
interplay between the exchange and the Zeeman terms in
Eq.~\eqref{e.2DQHAF}. The applied field breaks the $O(3)$ symmetry of
the isotropic model and induces a uniform alignment in the $z$
direction; such alignment frustrates the antiferromagnetic order
along $z$ but does not clash with antialignment on the $xy$ plane,
where $O(2)$ symmetry stays untouched. For infinitesimally small fields
one hence expects the spins to lay antialigned on the $xy$ plane, and
progressively cant out of it as $h$ is increased. Saturation occurs at
the critical value $h_{\rm{c}}=8$, above which the ground state
displays uniform ferromagnetic alignment along the $z$ direction. In
the range $0\,{<}\,h\,{<}\,h_{\rm{c}}$ one may also expect thermal
fluctuations of the $z$ spin components to be smaller for larger 
field, while no such a reduction should occur as far as the $x$ and $y$
components are concerned. The above picture clearly suggests the model
to share essential features with the easy-plane
2D\,QHAF~\cite{CRTVVprb2002}.

In the classical limit ($S\,{\to}\,\infty$, $JS^2\,{\to}\,J_{\rm{cl}}$,
$g\mu_{_{\rm B}}H\,{\to}\,h\,J_{\rm{cl}}$) both
analytical~\cite{VillainL1977,Pires1994} and
numerical~\cite{LandauB1981} calculations revealed the occurrence of a
Berezinskii-Kosterlitz-Thouless (BKT)
transition~\cite{Berezinskii1970,KosterlitzT1973}, for all values of
$h$ below saturation. In the quantum $S\,{=}\,\frac12$ case, evidence
of a field induced BKT transition was recently achieved for small
fields by means of quantum Monte Carlo (QMC) simulations based on the
continuous-time loop algorithm~\cite{TroyerS1998}. Unfortunately the
loop algorithm loses its efficiency exponentially as the field and/or
the inverse temperature are increased~\cite{OnishiNKM1999}: this fact
has so far prevented from a systematic investigation of the
strong-field regime~\cite{Syljuasen2000}. However, the recently
proposed {\it{worm}}~\cite{CRTVVprb2002} (or
{\it{directed-loop}}~\cite{SyljuasenS2002}) algorithm , working in any
lattice dimension, is effective also in the presence of a uniform field
of arbitrary intensity. Our QMC simulations are in fact based on such
an algorithm, which is a pure-quantum cluster algorithm (by this
meaning that it has no classical analogue, so far), that takes the
field into account by the dynamical nature of the process of cluster
(worm) growth, with the possibility that the worm traces back its route
after {\it bouncing} at some point. In particular the update process is
irreversible (i.e., the inverse of a single update step with finite
probability can have a vanishing probability), thus reflecting the
time-reversal symmetry breaking due to the presence of the field. A
detailed description of the algorithm is beyond the scope of this paper
and can be found in Ref.~\onlinecite{Roscilde2002}. Our simulations
were performed on a $L\,{\times}\,L$ square lattice ($L=16$, $32$,
$64$, and~$96$), each consisting of $10^4$ MC steps for thermalization
and of ($1$-$1.5$)$\,\times{10^5}$ MC steps for evaluation of
thermodynamic observables. During thermalization, the number of worms
produced at each step is adjusted so that the total length of the worms
in the imaginary-time direction roughly equals the size of the
($2{+}1$)-dimensional lattice, $L^2/t$; this number is then kept fixed
during the measurement phase. In this way, autocorrelation times of the
order of unity are achieved for all values of the field.

Thanks to this very effective tool, we got access to the thermodynamic
behavior of the model~\eqref{e.2DQHAF} with $S\,{=}\,\frac12$ and $h$
varying from $0$ to $h_{\rm c}$, looking for signatures of
field-induced BKT behavior that can be object of experimental
observation. In particular, we have focused our attention on the
specific heat $c(t,h)$ and the field-induced uniform magnetization
$m^z_{\rm{u}}(t,h)$, which are easily accessible to experiments, as
well as on the staggered magnetization along the field axis
$m^z_{{\rm{s}},L}(t,h)$, which provides further insight in the
microscopic ordering mechanism. Seven values of the field have been
considered: $h\,{=}\,0.1$, $0.2$, $0.4$, $1$, $2$, $4$, and~$6$.

\begin{figure}
\includegraphics[bbllx=1mm,bblly=11mm,bburx=181mm,bbury=154mm,%
     width=80mm,angle=0]{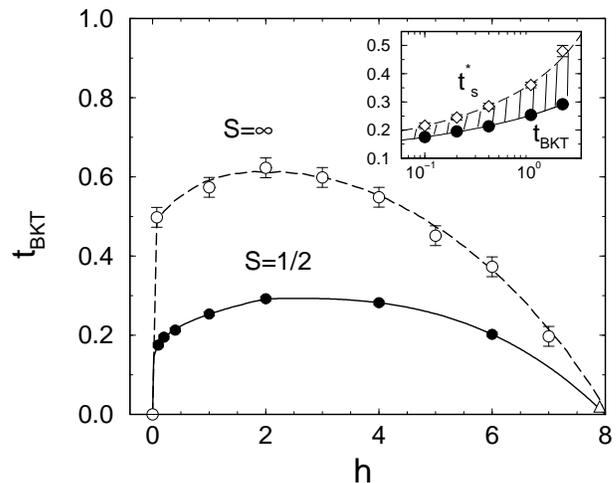}
\caption{\label{f.phdiagr} Phase diagram of the $S\,{=}\,\frac12$
2D\,QHAF in a magnetic field. Open symbols refer to the classical limit
of the model, from Ref.~\onlinecite{LandauB1981}; the triangle is a QMC
result from Ref.~\onlinecite{Syljuasen2000}. Inset:
$t_{_{\rm{BKT}}}(h)$ and $t^*_{\rm{s}}(h)$ (see text) for weak fields;
solid and dashed lines are logarithmic fits to the first three points
of each data set, and the shaded area marks the region of disordered
$XY$ behavior.}
\end{figure}

First of all we have performed a detailed finite-size scaling analysis
in order to check whether or not the predicted critical scaling
behavior of the in-plane staggered
susceptibility~\cite{KosterlitzT1973} and of the helicity
modulus~\cite{OlssonM1991} are reproduced at consistent temperatures:
having got a positive answer, we may state that a finite-temperature
phase transition of BKT type occurs in the model for all the considered
field values. The estimate of the critical temperature corresponding to
each field has been obtained via the same procedure used in
Ref.~\onlinecite{CRTVVprb2002} for the easy-plane 2D\,QHAF: the
resulting values are listed in Table~\ref{t.tBKTh}; they are consistent
with previous results~\cite{TroyerS1998} for $h\,{=}\,0.2$ and
$h\,{=}\,0.4$. In Fig.~\ref{f.phdiagr} we report the phase-diagram of
the model for $S\,{=}\,\frac12$ and $S=\infty$, the latter as from
classical MC results~\cite{LandauB1981}. We observe that the effect of
quantum fluctuations is limited to a strong renormalization of
$t_{_{\rm{BKT}}}$ with respect to the classical case, but the field
dependence is qualitatively the same.

\begin{table}
\caption{$t_{_{\rm{BKT}}}(h)$ as obtained by finite-size scaling
analysis.}
\label{t.tBKTh}
\begin{ruledtabular}
\begin{tabular}{cclcclcc}
 ~~~$h$~~~ & $t_{_{\rm{BKT}}}$ & ~~ &
 ~~~$h$~~~ & $t_{_{\rm{BKT}}}$ & ~~ &
 ~~~$h$~~~ & $t_{_{\rm{BKT}}}$ \\
\hline
 0.1 & 0.175(5) & & 1.0 & 0.254(5) & & 4.0 & 0.282(5) \\
 0.2 & 0.195(5) & & 2.0 & 0.292(5) & & 6.0 & 0.202(5) \\
 0.4 & 0.213(5) & &     &          & &                \\
\end{tabular}
\end{ruledtabular}
\end{table}

Let us first concentrate on the weak-field regime: from various
analytical arguments~\cite{Okwamoto1984,Pires1994,ZhitomirskyN1998} one
can infer a mapping of the 2D\,QHAF in a field on a weakly easy-plane
magnet with exchange anisotropy, i.e., Hamiltonian~\eqref{e.2DQHAF}
with $H=0$ and the additional term
$-J\Delta\sum_{\bm{i},\bm{d}}\hat{S}^z_{\bm{i}}\hat{S}^z_{\bm{i}+\bm{d}}$;
the effective exchange anisotropy $\Delta$ is expected to scale with
the field as $\Delta\,{\propto}\,h^2$. Hence, the BKT critical
temperature is expected to obey, both in the classical and in the
quantum case~\cite{CRTVVprb2002}, the following expression
\begin{equation}
 t_{_{\rm{BKT}}}(h) \simeq \frac{4\pi\rho{_{_S}}/J}{\ln (C/h^2)}
\label{e.tbkth}
\end{equation}
where $\rho{_{_S}}$ is the spin stiffness of the isotropic model, and
$C$ a constant. Fitting our results for the lowest three values of $h$
to the above expression we get $\rho_{_S}\,{\simeq}\,0.22\,J$, and
excluding the third value we find $\rho_{_S}\,{\simeq}\,0.19$.
Remarkably, these values are consistent with the renormalized spin
stiffness of the $S\,{=}\,\frac12$ isotropic 2D\,QHAF,
$\rho_{_S}\,{=}\,0.180~J$ \cite{BeardBGW1998}. We also
notice that the fitting curve keeps interpolating the data up to
$h\,{\lesssim}\,2$. Finally, by directly comparing the low-field
behavior of $t_{_{\rm{BKT}}}(h)$ with that of $t_{_{\rm BKT}}(\Delta)$
in the weakly planar antiferromagnet~\cite{CRTVVprb2002}, we obtain an
excellent agreement for $\Delta\approx(3/8)\,h^2$.

Moving towards higher fields two different effects are expected:
({\it{i}}\,) the fluctuations of the $z$ components become smaller,
resulting in an enhanced effective easy-plane anisotropy, and
({\it{ii}}\,) the average projection of the spins in the $xy$ plane
decreases, due to the increasing uniform magnetization. Globally, the
system behaves as a {\it renormalized} planar rotator with
progressively {\it reduced} rotator length. It is remarkable that,
despite the spin configurations being characterized by a smaller and
smaller projection in the $xy$ plane, the $XY$ character is apparent
even for $h$ close to $h_{\rm{c}}$ and the transition verifies all the
predictions of the BKT theory. The interplay between the two competing
field effects yields the non-monotonous dependence of
$t_{_{\rm{BKT}}}(h)$: for low field the reduction of $z$ fluctuations
is dominant and $t_{_{\rm{BKT}}}(h)$ increases with $h$, starting as in
Eq.~\eqref{e.tbkth}, while for higher field spin canting prevails and
$t_{_{\rm{BKT}}}(h)$ decreases, eventually vanishing at the saturation
field $h_{\rm{c}}$. Therefore, a maximum in $t_{_{\rm{BKT}}}(h)$
connects the two limiting behaviors, as already observed in the
classical phase diagram: the comparison in Fig.~\ref{f.phdiagr} shows
that for $S\,{=}\,\frac12$ the maximum shifts to slightly higher field
as a consequence of quantum fluctuations.

\begin{figure}
\includegraphics[bbllx=25mm,bblly=142mm,bburx=200mm,bbury=269mm,%
     width=86mm,angle=0]{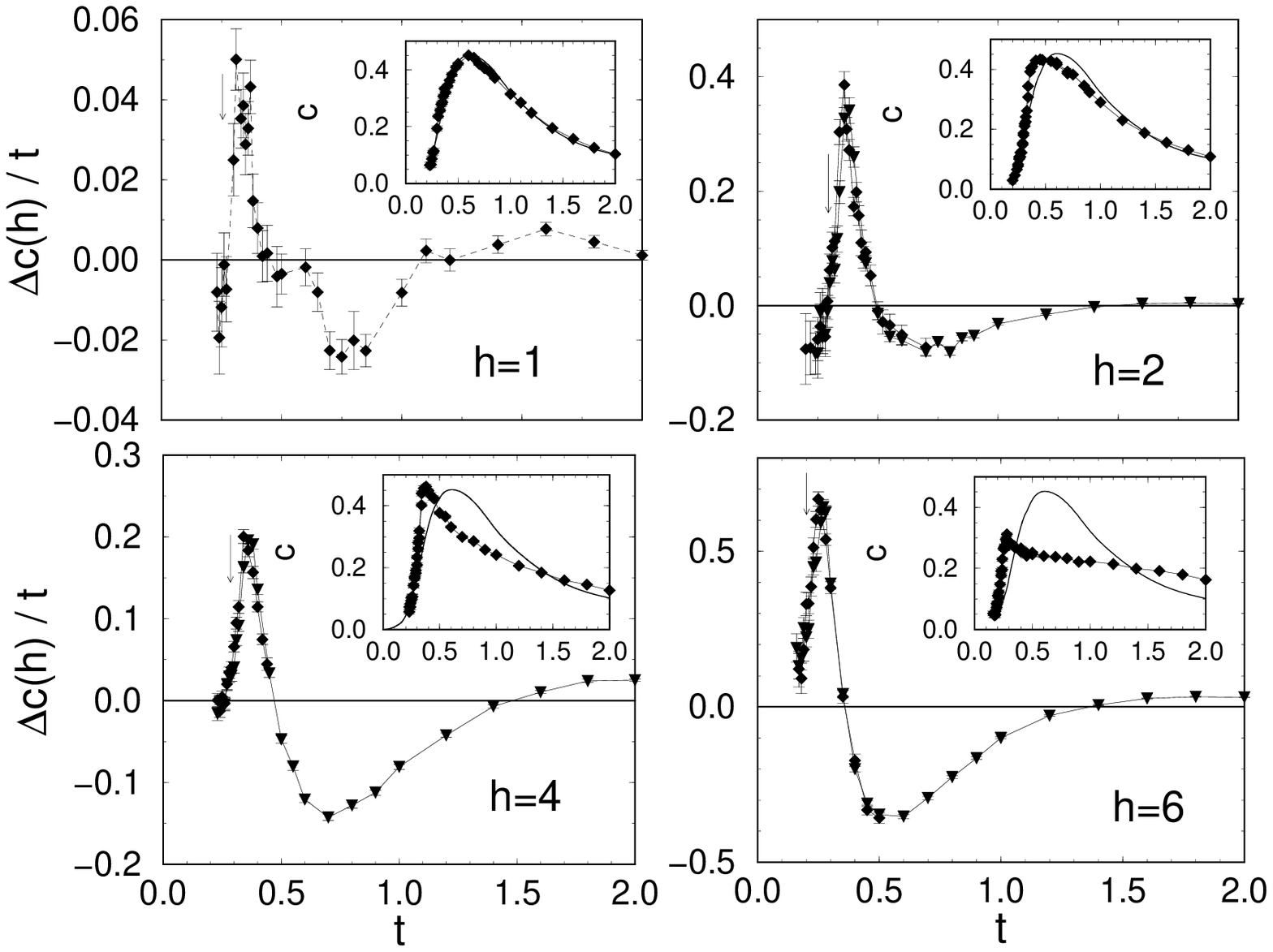} 
\caption{\label{f.sph} $\Delta{c}(t,h)/t=[c(t,h)-c(t,0)]/t$ vs
temperature, for four different field values. Insets: magnetic specific
heat $c(t,h)$ compared to $c(t,0)$ (thick solid line). The zero-field
specific heat $c(t,0)$ is obtained by interpolating numerical and
analytical data from Refs.~\onlinecite{MakivicD1991,KimT1998,
Takahashi1989}); the arrows mark the estimated BKT critical
temperatures.}
\end{figure}

We now consider the temperature dependence of some relevant
observables, beginning with the specific heat $c(t,h)$. While our data
resolution for $h\,{\lesssim}\,0.4$ prevents from observing
significant deviations with respect to the zero-field system, the
results for the four largest fields are shown in Fig.~\ref{f.sph}; in
particular, what is plotted is the specific heat variation upon
application of the field, $\Delta{c}(t,h)=[c(t,h)-c(t,0)]$, divided by
$t$. This quantity equals the difference of the entropy derivatives
$\partial_tS(t,h)-\partial_tS(t,0)$ and allows us to draw the
following picture: At low temperature the entropy increase is smaller
than in zero field reflecting the presence of quasi long-range order
induced by the field via stabilization of bound vortex/antivortex
pairs. Slightly above the BKT transition ($t$ about
$20\,{\div}\,30\,\%$ larger) a sharp entropy increase occurs, which we
interprete as due to vortex unbinding. When the temperature is further
raised all vortices are free and then $S(t,h)$ increases slower than
$S(t,0)$; eventually the entropy difference vanishes in the fully
disordered system at $t\,{\to}\,\infty$. For incresing field the peak
of $c(t,h)$, which mimics the BKT peak of the $XY$ model, moves to
lower temperature, thereby getting narrower, as shown in the insets of
Fig.~\ref{f.sph}. It is worth noting that $\Delta{c}(t,h)$, which
bears very clear signatures of the BKT behavior, is also easily
accessible to experiments, because non-magnetic contributions cancel
in its definition.

\begin{figure}
\includegraphics[bbllx=3mm,bblly=13mm,bburx=198mm,bbury=146mm,%
     width=80mm,angle=0]{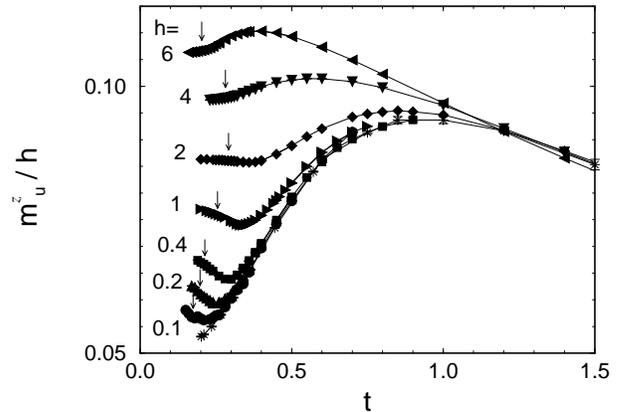}
 \caption{\label{f.szu} Field-induced uniform magnetization
$m^z_{\rm{u}}(t,h)$ vs temperature, for different field values. The
stars represent the zero-field uniform susceptibility from
Refs.~\onlinecite{MakivicD1991,KimT1998}; arrows as in
Fig.~\ref{f.sph}.}
\end{figure}

We have also considered the field-induced uniform magnetization,
$m^z_{\rm{u}}(t,h)\equiv\langle\hat{S}^z_{\bm{i}}\rangle$; such a
quantity, which can be experimentally determined via standard
magnetometry measurements, is also a highly precise output of the QMC
simulations. In Fig.~\ref{f.szu} we report $m^z_{\rm{u}}(t,h)/h$ for
different fields. For $h\,{\lesssim}\,2$ and for high enough
temperature this quantity is found to coincide with the uniform
susceptibility of the zero-field case, thus showing that the
magnetization process is linear. Upon lowering $t$ the nonlinearities show
up and the uniform magnetization changes completely its temperature
dependence, displaying a minimum at $t=t^*_{\rm{u}}(h)$. This feature
marks the onset of $XY$ behavior: as the temperature is lowered below
$t^*_{\rm{u}}$, the system is increasingly magnetized along the
$z$-axis and the short-range antiferromagnetic correlations of the $z$
spin components, as well as their thermal fluctuations, are suppressed,
in turn stabilizing the canted configurations. A remarkable feature is
that this crossover to $XY$ behavior is located at a temperature
$t^*_{\rm{u}}(h)$ well above the critical point. For $h\,{\gtrsim}\,2$
the minimum above $t_{_{\rm{BKT}}}$ disappears, and the most prominent
feature is rather the shift of the broad maximum to lower temperature.

\begin{figure}
\includegraphics[bbllx=3mm,bblly=11mm,bburx=203mm,bbury=166mm,%
     width=80mm,angle=0]{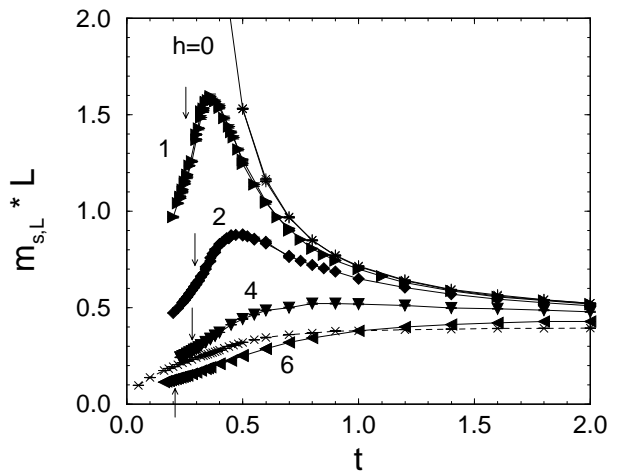}
 \caption{\label{f.szstaggL} Finite-size staggered magnetization
$m^z_{{\rm{s}},L}(t,h)$ vs temperature, for different field values. The
$\times$'s represent the same quantity for the $S\,{=}\,\frac12$
2D\,$XY$ model. Arrows as in Fig.~\ref{f.sph}.}
\end{figure}

The crossover from isotropic to $XY$ behavior in low fields
($h\,{\lesssim}\,2$) can be also detected in the temperature dependence
of the finite-size staggered magnetization along the hard
$z$-axis~\cite{CRVV2003},
$m^z_{{\rm{s}},L}(t,h)\equiv L^{-2}\big\langle\big|\sum_{\bm{i}}(-)^{\bm{i}}
\hat{S}_{\bm{i}}^z\big|\big\rangle$, shown in Fig.~\ref{f.szstaggL}.
In absence of long-range order $m^z_{{\rm{s}},L}$ is known to scale to
zero as $1/L$ (for large enough $L$)~\cite{BinderH1988}, and
$m^z_{{\rm{s}},L}L$ is hence a bulk property of the system. In the
limit $t\to\infty$ the system behaves as a collection of paramagnetic
spins so that, by applying the central-limit theorem, one finds
$m^z_{{\rm{s}},L}~L\to\sqrt{2S(S+1)/(3\pi)}\approx 0.399$ for
$S\,{=}\,\frac12$. As $t$ decreases, $m^z_{{\rm{s}},L}~L$ in the
isotropic 2D\,QHAF monotonically increases and diverges for
$t\,{\to}\,0$. In the 2D $XY$ model the same quantity decreases below
the infinite-$t$ value, due to the suppression of out-of-plane
fluctuations. The coexistence of both the above behaviors is most
clearly observed for $h\,{\lesssim}\,2$, and the appearance of a
maximum in $m^z_{{\rm{s}},L}$ at $t\,{\equiv}\,t^*_{\rm{s}}$ also marks
the crossover from isotropic to $XY$ behavior.

Hitherto, we have identified two ways of locating a crossover
temperature, at least for low and intermediate fields. One can check
that for low fields these temperatures match with each other,
$t^*_{\rm{s}}(h)\,{\simeq}\,t^*_{\rm{u}}(h)$, so that the crossover
temperature is unambiguous and its estimates are
$t^*_{\rm{s}}(0.1)\,{=}\,0.22(1)$, $t^*_{\rm{s}}(0.2)\,{=}\,0.25(1)$
and $t^*_{\rm{s}}(0.4)\,{=}\,0.29(1)$. In this regime the crossover
temperature is expected to follow a logarithmic behavior analogous to
that of Eq.~\eqref{e.tbkth} (with a different coefficient $C'$) as
suggested in Refs.~\onlinecite{CRTVVprb2002} and~\onlinecite{CRVV2003};
a fit of $t^*_{\rm{s}}(h)$ for $h\,{\leq}\,0.4$ gives $\rho_{_S}=0.19$,
again in good agreement with the known value~\cite{BeardBGW1998} of the
spin stiffness for $S\,{=}\,\frac12$. For larger fields
$t^*_{\rm{s}}(h)$ is systematically higher than $t^*_{\rm{u}}(h)$,
suggesting that the crossover phenomenon extends over a wider
temperature range, but we note that for intermediate fields the same
fitting function still interpolates the data,
$t^*_{\rm{s}}(1)\,{=}\,0.36(1)$ and $t^*_{\rm{s}}(2)=0.48(2)$, as shown
in the inset of Fig.~\ref{f.phdiagr}. Finally, for strong fields the
explicit signature of the crossover gradually disappears from
$m^z_{{\rm{s}},L}*L$, and for $h\,{=}\,6$ this quantity is nearly
monotonic as in the $XY$ model; the antiferromagnetic interaction of
the $z$ components is then almost completely overcome by the applied
field.

In conclusion, we have studied the $S\,{=}\,\frac12$
two-di\-men\-sional quantum Heisenberg antiferromagnet on the square
lattice in an arbitrary uniform field by means of the quantum Monte
Carlo method based on the worm algorithm. Our results point out that an
arbitrarily small field is able to induce a BKT transition and an
extended $XY$ phase above it, as in the case of an easy-plane exchange
anisotropy. The field-induced $XY$ behavior becomes more and more
marked for increasing fields, while for strong fields the
antiferromagnetic behavior along the field axis is nearly washed out,
so that the system behaves as a planar rotator model with
antiferromagnetism surviving in the orthogonal plane only; the BKT
critical temperature vanishes as the field reaches the saturation value
$h_{\rm{c}}$ and the effective rotator length goes to zero. The
model in a moderately strong field represents an ideal realization of
the $XY$ model: $XY$ behavior can be detected by measuring standard
non-critical quantities, as the specific heat or the induced
magnetization; this opens the possibility for an experimental
realization of the $XY$ model in purely magnetic systems, and for a
systematic investigation of the dynamics of vortex/antivortex
excitations.

Useful discussions with V.~Tognetti, A.~Rigamonti, and P.~Carretta are
gratefully acknowledged . This work has been partially performed on the
parallel beowulf cluster at CINECA (Bologna, Italy) through INFM grant
1031154758.

\end{document}